\documentclass[aps,preprint,superscriptaddress,11pt]{revtex4}

\usepackage{amsmath}
\usepackage{graphics}
\usepackage{graphicx}
\usepackage[mathcal]{eucal}

\begin{document}



\preprint{FERMILAB-PUB-06-346-T, COLO-HEP-518, UCI-TR-2006-16}

\title{Low Scale Seesaw, Electron EDM and Leptogenesis in a Model with Spontaneous CP Violation}


\author{Mu-Chun Chen}
\email[]{mcchen@fnal.gov}
\affiliation{Theoretical Physics Department, Fermilab, Batavia, IL 60510-0500, USA\\
and}
\affiliation{Department of Physics \& Astronomy, University of California, Irvine, CA 92697, USA}
\author{K.T. Mahanthappa}
\email[]{ktm@pizero.colorado.edu}
\affiliation{Department of Physics, University of Colorado at Boulder, Boulder, CO 80309-0390, USA}



\begin{abstract}
Strong correlations between leptogenesis and low energy CP violating leptonic processes have been shown by us to exist in the minimal left-right symmetric model with spontaneous CP violation. In this note, we investigate the implications of this model for the electric dipole moment of the electron. With an additional broken $U(1)_{H}$ symmetry, the seesaw scale can be lowered to close to the electroweak scale. This additional symmetry also makes the connection between CP violation in quark sector to that in the lepton sector possible.
\end{abstract}


\maketitle


\section{Introduction}

In recent years, many attempts have been made~\cite{Chen:2004ww,Joshipura:2001ui,Branco:2003rt} 
to find the connections between leptogenesis~\cite{Fukugita:1986hr} 
and low energy CP violating processes.  
In Ref.~\cite{Chen:2004ww}, we found that a strong correlation among CP violation in leptogenesis, neutrinoless double beta decay and neutrino oscillation exists in minimal left-right symmetric model~\cite{Pati:1974yy} 
assuming CP violation occurs spontaneously. In the Standard Model (SM) with massless neutrinos, lepton EDM can arise only at the four loop order and thus the prediction is highly suppressed. (For review, see, for example, Ref.~\cite{Bernreuther:1990jx}.)  For the electron, the SM prediction is $d_{e} \sim 10^{-38}$ e-cm, which is ten orders of magnitude smaller compared to the current bound  
$d_{e} < 1.6 \times 10^{-27}$ e-cm at 90\% confidence level~\cite{Regan:2002ta}. 
By having massive neutrinos in the SM, a slightly enhanced value of $d_{e} \sim 10^{-33}$ e-cm can be induced at the two loop order with fine-tuned model  parameters. Due to the existence of the $SU(2)_{R}$ gauge boson, $W_{R}$, non-vanishing lepton EDM can appear in the LR model even at the one-loop order. The predicted values for $d_{e}$ can therefore be accessible experimentally in the LR model, provided that the $SU(2)_{R}$ breaking scale is only a few orders of magnitude above the EW scale.  Such a scale can be obtained by introducing a broken $U(1)_{H}$ symmetry~\cite{Khasanov:2001tu}. In the minimal version of the LR model with spontaneous CP violation, the size of the electron EDM is related to other CP violating observable as there are {\it only} two intrinsic CP violating phases. In this paper, we consider a model based on left-right symmetry, in which CP violation occurs spontaneously, in addition to having a low seesaw scale.  We investigate the implications of this model for neutrino oscillation, leptogenesis and the electric dipole moment (EDM) of the electron and demonstrate the strong correlation among CP violation in these processes. This paper is organized as follows. In Sec.~\ref{sec:model}, we describe our model and show how spontaneous CP violation can occur and how seesaw scale can be lowered. This is followed by Sec.~\ref{sec:edm} in which the calculation of the electron EDM in this model is given. Our results are presented in Sec.~\ref{sec:result}. Sec.~\ref{sec:conclude} concludes this note.

\section{The Model}\label{sec:model}

The minimal left-right symmetric model is based on the gauge group $SU(2)_{L}\times SU(2)_{R} \times U(1)_{B-L}\times P$, where the parity $P$ acts on the two $SU(2)$ groups. 
The Higgs sector contains one bi-doublet, $\Phi$, one $SU(2)_{L}$ triplet, $\Delta_{L}$, and one $SU(2)_{R}$ triplet, $\Delta_{R}$. 
In general, there are four CP violating phases associated with the VEV's in the neutral components of these Higgs fields.  
Nevertheless, using the global degrees of freedom, these four CP violating phases can be rotated away, reducing the number to two, 
$\alpha_{\kappa^{\prime}}$ and $\alpha_{L}$, as chosen in the VEV of the three fields,
\begin{eqnarray}
\Phi  =  \left(
\begin{array}{cc}
\kappa & 0 \\
0 & \kappa^{\prime} e^{i\alpha_{\kappa^{\prime}}}
\end{array}\right) \;, \quad
\Delta_{L}  =  \left(\begin{array}{cc}
0 & 0\\
v_{L}e^{i\alpha_{L}} & 0
\end{array}\right) \; , \quad
\Delta_{R} =  \left(\begin{array}{cc}
0 & 0\\
v_{R} & 0
\end{array}\right) \; ,
\end{eqnarray}
where $v^{2} = |\kappa|^{2} +| \kappa^{\prime}|^{2} \simeq 2M_{W}^{2}/g^{2} \simeq (174 \; \mbox{GeV})^{2}$ with $M_{W}$ being the mass of the $W$ gauge boson and $g$ being the $SU(2)_{L}$ gauge coupling constant.   
To satisfy the constraints on flavor changing neutral currents, the ratio of the bi-doublet VEV's, $r \equiv \kappa^{\prime}/\kappa$, has to be $r \lesssim m_{b}/m_{t}$. 

The $SU(2)_{R}$ breaking scale can be significantly lowered if an additional broken $U(1)_{H}$  symmetry is imposed in the lepton sector~\cite{Khasanov:2001tu}. The $U(1)_{H}$ symmetry is broken by the VEV of a scalar field $S$, which has $U(1)_{H}$ charge $Q(S) = +1$ and is a singlet under the LR symmetry group. Due to the $U(1)_{H}$ symmetry, many of the interactions can now appear only as higher dimensional non-renormalizable operators that couple to the scalar field $S$. After integrating out the field $S$, these interactions are suppressed by some powers of a small parameter $\omega \equiv \left< S \right> / M_{S}$, where $M_{S}$ is the mass of the field $S$. We indicate the $U(1)_{H}$ charge of the field $f$ as $Q(f)$, where $f = (\Delta_{L}, \; \Delta_{R}, \; \Phi, \; \ell_{L}, \; \ell_{R}, \; q_{L}, \; q_{R})$ with $\ell_{L}$ ($q_{L}$) and $\ell_{R}$ ($q_{R}$) being left-handed and right-handed lepton (quark) doublets, respectively. To illustrate the dependence on the mixing angle, we further assume in this paper that the $U(1)_{H}$ charges are the same among all three generations of fermions. Thus this $U(1)_{H}$ symmetry is crucial only for lowering the seesaw scale, but irrelevant for the observed flavor mixing (For a review on how the observed mixing angles can be accounted for with a horizontal symmetry, see e.g. Ref.~\cite{Chen:2003zv}.)  
The scalar potential of the model as well as the minimization conditions are given in the Appendix.  
Due to the presence of the broken $U(1)_{H}$ symmetry, the following parameters in the scalar potential are suppressed,
\begin{eqnarray}
(\mu_{2}^{2}, \, \lambda_{4}, \; \alpha_{2}) & \rightarrow & \omega^{2|Q(\Phi)|} (\mu_{2}^{2}, \, \lambda_{4}, \; \alpha_{2}) \; , \\
\lambda_{2} & \rightarrow  & \omega^{4|Q(\Phi)|} \lambda_{2} \; ,
\\ 
\beta_{1} & \rightarrow & \omega^{|Q(\Delta_{R}) - Q(\Delta_{L})|} \beta_{1} \; ,\\
\beta_{2} & \rightarrow & \omega^{|Q(\Delta_{R}) - Q(\Delta_{L})-2Q(\Phi)|} \beta_{2} \; , \\
\beta_{3} & \rightarrow & \omega^{|Q(\Delta_{R}) - Q(\Delta_{L})+2Q(\Phi)|} \beta_{3} \; , \\
\rho_{4} & \rightarrow & \omega^{2|Q(\Delta_{L})-Q(\Delta_{R})|} \rho_{4} \; .
\end{eqnarray}
The rest of the parameters in the scalar potential are not suppressed, since the interactions due to these couplings are allowed by the $U(1)_{H}$ symmetry. The seesaw relation between the two triplet VEV's, $v_{L}$ and $v_{R}$, is then given by,
\begin{equation}
v_{L} \simeq \beta \frac{\kappa^{2}}{v_{R}} \; ,
\end{equation}
where the parameter $\beta$ is,
\begin{equation}
\beta \sim \beta^{\prime} \cdot \mbox{Max} \big\{
\omega^{|Q(\Delta_{R} )- Q(\Delta_{L}) - 2 Q(\Phi)|}, \, 
r\omega^{|Q(\Delta_{R} )- Q(\Delta_{L})|}, \, 
r^{2} \omega^{|Q(\Delta_{R} )- Q(\Delta_{L}) + 2 Q(\Phi)|} \big\}
\; ,
\end{equation}
followed directly from Eq.~\ref{eq:minimize1}. The parameter $\beta^{\prime}$ in the equation above is a function of $\mathcal{O}(1)$ parameters $(\beta_{1,2,3}, \; \rho_{1,3}$) in the scalar potential.  

From Eq.~\ref{eq:minimize3}, it can be seen immediately that the phase $\alpha_{\kappa^{\prime}}$ scales as $\sin \alpha_{\kappa^{\prime}} \sim  (\beta/\alpha_{3})(v_{L}/v_{R})$.  
Because the LR breaking scale $v_{R}$ is much lowered, it may appear at the first glance that the presence of this additional $U(1)_{H}$ symmetry could alleviate the fine-tuning that is 
needed  in order to generate a sizable value for $\alpha_{\kappa^{\prime}}$ as pointed out previously~\cite {Deshpande:1990ip}  . This turns out not to be the case as the proportionality constant between $\alpha_{\kappa^{\prime}}$ and $v_{L}/v_{R}$ is also suppressed.  
We emphasize, however, that there is no such fine-tuning needed in order to generate a large phase $\alpha_{L}$ in the lepton sector, regardless of the scale of $v_{R}$~\cite{Rodriguez:2002ey}. The fine-tuning can be alleviated by having extra dimensions in which the seesaw scale $v_{R}$ is lowered to the electroweak scale due to the warped geometry~\cite{Chen:2005mz}, or by having an additional singlet scalar field~\cite{Branco:2006av}. We leave these possibilities for further investigation in a later paper.

The Yukawa interactions in the lepton sector are described by,
\begin{equation}
\mathcal{L}_{Yuk}^{\ell} = 
\overline{L}_{i,L} (P_{ij} \Phi + R_{ij} \widetilde{\Phi}) L_{j,R} 
+ i f_{ij} ( L_{i,L}^{T} C \tau_{2} \Delta_{L} L_{j,L} + L_{i,R}^{T} C\tau_{2} \Delta_{R} L_{j,R}) + h.c.
\end{equation} 
Due to the $U(1)_{H}$ symmetry, various mass matrices in the lepton sector are also suppressed by powers of $\omega$. 
The charged lepton mass matrix now reads,
\begin{eqnarray}
M_{e} &=  P_{ij} \kappa^{\prime} e^{i\alpha_{\kappa^{\prime}}} \omega^{|Q(L_{L})-Q(L_{R}) - Q(\Phi)|} 
 + R_{ij} \kappa \omega^{|Q(L_{L})-Q(L_{R}) + Q(\Phi)|} 
\end{eqnarray}
and the neutrino Dirac mass matrix becomes,
\begin{eqnarray}
M_{\nu_{D}}  =  P_{ij} \kappa  \omega^{|Q(L_{L})-Q(L_{R}) - Q(\Phi)|} 
+ R_{ij} \kappa^{\prime} e^{-i\alpha_{\kappa^{\prime}}} \omega^{|Q(L_{L})-Q(L_{R}) + Q(\Phi)|} \; . \label{eq:nuDirac}
\end{eqnarray}
Due to the presence of the left-right parity,  a Majorana mass term for the LH neutrinos must be present. The two Majorana mass terms are proportional to one another and are given by,
\begin{equation}
M_{\nu}^{RR} = f_{ij} v_{R} \omega^{|Q(\Delta_{R})+ 2Q(L_{R})|}\; , \quad
M_{\nu}^{LL} = f_{ij} v_{L} e^{i\alpha_{L}} \omega^{|Q(\Delta_{L}) + 2Q(L_{L})|} \; . \label{eq:nuMaj}
\end{equation}
Similarly, the Yukawa interactions in the quark sector,
\begin{equation}
\mathcal{L}_{Yuk}^{q} = \overline{Q}_{i,L} ( F_{ij} \Phi + G_{ij} \widetilde{\Phi}) Q_{j,R} \; ,
\end{equation}
lead to the following quark mass matrices, 
\begin{eqnarray}
M_{u} & = &  F_{ij} \kappa  \omega^{|Q(q_{L})-Q(q_{R}) - Q(\Phi)|} 
+ G_{ij} \kappa^{\prime} e^{i\alpha_{\kappa^{\prime}}} \omega^{|Q(q_{L})-Q(q_{R}) + Q(\Phi)|} 
\\
M_{d} & = & F_{ij} \kappa^{\prime} e^{i\alpha_{\kappa^{\prime}}} \omega^{|Q(q_{L})-Q(q_{R}) - Q(\Phi)|} 
 + G_{ij} \kappa \omega^{|Q(q_{L})-Q(q_{R}) + Q(\Phi)|}  \; .
\end{eqnarray}
Depending on the choices for the quark charges, $Q(q_{L})$ and $Q(q_{R})$, the quark masses can be suppresses by powers of $\omega$, as in the lepton sector. This can then be utilized to explain the hierarchy between $m_{b}$ and $m_{t}$. The CKM matrix can then be obtained by diagonalizing both mass matrices for the up-type quarks and the down-type quarks, and all CP violating processes in the quark sector can then be explained by a single phase, $\alpha_{\kappa^{\prime}}$. It has been pointed out, however, that CP violation caused by $\alpha_{\kappa^{\prime}}$ alone, is not sufficient to explain to observed asymmetry in the decay of   $B \rightarrow \psi K_{s}$~\cite{Bergmann:2001pm}. 
In our discussion below, only the Yukawa interactions in the lepton sector are relevant. The consequences of having non-trivial quark charges as well as how sufficient CP violation in B decay can be obtained  will be investigated in a subsequent paper.

\section{Electric Dipole Moment of the Electron}
\label{sec:edm}

The EDM of the leptons in left-right symmetric model was first calculated in Ref.~\cite{Ecker:1983dj}. Here we briefly summarize the result. In the minimal LR model, the charged current interaction in the lepton sector can be mediated by both $W_{L}$ and $W_{R}$ gauge bosons, and it is given by,
\begin{equation}
\mathcal{L}_{cc}  =  -\frac{g}{\sqrt{2}} \sum_{i} \left( \overline{L}_{i_{L}} \gamma^{\mu} \nu_{i_{L}} W_{L,\mu}^{-}+ \overline{L}_{i_{R}} \gamma^{\mu} N_{i_{R}} W_{R,\mu}^{-} \right) 
 + h.c. \; .
\end{equation}
Upon electroweak symmetry breaking, the mass matrix of the $W$ gauge bosons is,
\begin{equation}
\left( \begin{array}{cc} W_{L}^{+} &  W_{R}^{+} \end{array} \right)
\left(\begin{array}{ccc}
\frac{g^{2}}{2}(2v_{L}^{2} + \kappa^{2} + \kappa^{\prime 2} ) & \;\; & -g^{2} \kappa\kappa^{\prime} e^{i\alpha_{\kappa^{\prime}}} \\
 -g^{2} \kappa\kappa^{\prime} e^{-i\alpha_{\kappa^{\prime}}} &  & \frac{g^{2}}{2} (2v_{R}^{2} + \kappa^{2} + \kappa^{\prime2})
\end{array}\right) 
\left(\begin{array}{c}
W_{L}^{-} \\ W_{R}^{-}
\end{array}\right)
 \; .
\end{equation}
The mass eigenstates $W_{1,2}$ are related to the flavor eigenstates $W_{L,R}$ by,
\begin{equation}
\left(\begin{array}{cc} W_{L}^{+} &  W_{R}^{+} \end{array} \right) = U \left(\begin{array}{c}
W_{1}^{+} \\ 
W_{2}^{+}
\end{array}\right) \; ,
\end{equation}
where the unitary matrix $U$  that diagonalize the $W$ boson mass matrix is given by,
\begin{equation}
U = \left(\begin{array}{cc}
\cos\xi & -\sin\xi \; e^{i\alpha_{\kappa^{\prime}}} \\
\sin\xi \;  e^{-i\alpha_{\kappa^{\prime}}} & \cos\xi
\end{array}\right) \; ,
\end{equation}
and the two mass eigenvalues obey $M_{1} \simeq M_{W} < M_{2}$. 
The left-right mixing angle, $\xi$, is suppressed by the left-right breaking scale $v_{R}$. It is give by,
\begin{equation}
\tan2\xi = \frac{2\kappa\kappa^{\prime}}{v_{R}^{2} - v_{L}^{2}}  \; .
\end{equation}

The seesaw matrix for the neutrinos now reads, 
\begin{equation}
\left(\begin{array}{cc} \overline{\nu}^{c} &  \overline{N} \end{array} \right)_{R} \left(\begin{array}{cc}
M_{\nu}^{LL} & M_{\nu_{D}}\\
M_{\nu_{D}}^{T} & M_{\nu}^{RR}
\end{array}\right)
\left(\begin{array}{c}
\nu \\
N^{c}
\end{array}\right)_{L} + h.c. \; .
\end{equation}
where the Dirac mass term, $M_{\nu_{D}}$, and the Majorana mass terms, $M_{\nu}^{LL}$ and $M_{\nu}^{RR}$, are give in Eq.~\ref{eq:nuDirac} and 
Eq.~\ref{eq:nuMaj} respectively. This seesaw mass matrix can be diagonalized by a $6 \times 6$ unitary matrix $V$, 
\begin{equation}
V = \left(\begin{array}{c}
V_{L} \\ V_{R}^{\ast} 
\end{array}\right) \; .
\end{equation}
The flavor eigenstates, $\nu$ and $N^{c}$, and the mass eigenstates, $\psi$, of the neutrinos are related by,
\begin{eqnarray}
\left(\begin{array}{c}
\nu \\ N^{c}
\end{array}\right) & = & V \psi_{L} \; ,
\\
\left( \overline{\nu}^{c}  \overline{N}\right)_{R} & = & \overline{\psi}_{R} V^{T} \; .
\end{eqnarray}
and thus the flavor eigenstates can be written in terms of the mass eigenstates and elements of the diagonalization matrix as,
\begin{eqnarray}
\nu_{L_{i}} & = & \sum_{j=1}^{6} (V_{L})_{ij}  \psi_{L_{j}} \; , \\
N_{L_{i}}^{c} & = & \sum_{j=1}^{6} (V_{R}^{\ast})_{ij} \psi_{L_{j}} \; .
\end{eqnarray}
With these definitions, the charged current interaction in the lepton sector can then be written in the mass eigenstates of the neutrinos and the gauge bosons as,
\begin{widetext}
\begin{equation}
\mathcal{L}_{cc}  =  -\frac{g}{\sqrt{2}} 
\sum_{i=1}^{n} \sum_{j=1}^{6} \sum_{a=1,2} 
W^{\mu}_{a} \left[ U_{La} V_{L_{ij}} \overline{L}_{i_{L}} \gamma_{\mu} \psi_{j_{L}} 
+ U_{Ra} V_{R_{ij}} \overline{L}_{i_{R}} \gamma_{\mu} \psi_{j_{R}} 
\right] + h.c. \quad . 
\end{equation}
\end{widetext}

\begin{figure}[t!]
\begin{tabular}{ccc}
\includegraphics{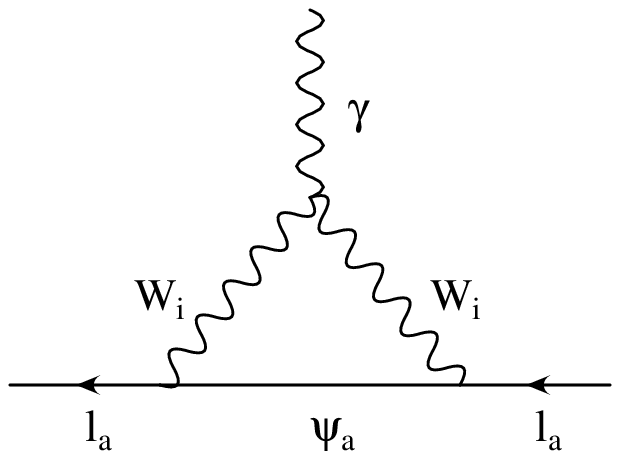} & \hspace{0.2in} & \includegraphics{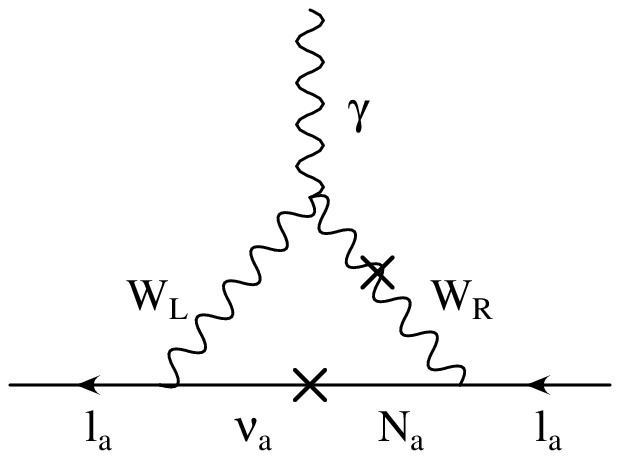}
\\
(a) & & (b)\\
\end{tabular}
\caption{One-loop diagram in minimal LR model that contribute to the electron EDM: (a) Diagram shown with particle in their mass eigenstates; (b) Diagram shown with particles in the weak charged current interaction eigenstates.}
\label{fig:edm}
\end{figure}

Non-vanishing EDM of the electron in the minimal LR model arises at one-loop from diagrams mediated by the $W$ bosons and the neutrinos, as shown in Fig.~\ref{fig:edm}. 
The total contribution is given by~\cite{Ecker:1983dj}, 
\begin{widetext}
\begin{equation}
d_{e} = \frac{eM_{W}^{2}G_{F}}{4\sqrt{2}\pi^{2}} \sum_{a=1,2} 
\biggl( \frac{U_{La} U_{Ra}}{M_{a}^{2}} \biggr) \sum_{j=1}^{6} m_{j}  \Im (V_{L_{1j}} V_{R_{1j}}^{\ast}) I_{1} \biggl( \frac{m_{j}^{2}}{M_{a}^{2}},\frac{m_{e}^{2}}{M_{a}^{2}}\biggr)
\; .
\end{equation} 
\end{widetext}
Here, $M_{a}$ (for $a=1,2$) are masses of the two $W$ boson masses, and $m_{j}$ are the neutrino masses. The loop integral $I_{1}(x,y)$ is given by,
\begin{eqnarray}
I_{1}(x,y) & = & \frac{1}{2} + 3F_{0}(x,y) - 6 F_{1}(x,y) + (3-y) F_{2}(x,y) + y F_{3}(x,y)
\\
& \simeq & 
\frac{2}{(1-x)^{2}} \biggl[ 1 - \frac{11}{4} x + \frac{1}{4} x^{2} - \frac{3x^{2} \ln x}{2(1-x)} \biggr] \; ,
\nonumber
\end{eqnarray}
where the function $F_{n}(x,y)$ (for $n=1,2,3$) is defined as, 
\begin{equation}
F_{n}(x,y) = \int_{0}^{1} dt \, \frac{t^{n}}{1-t-yt(1-t)+xt} \; .
\end{equation}
In the limit $y \rightarrow 0$, this function can be approximated as,
\begin{equation}
I_{1}(x_{ij},0) \simeq \bigg\{
\begin{array}{ll}
 1 + \frac{6\ln x_{i1}}{x_{i1}}  \; , & \mbox{for} \; m_{i} \ll M_{1} 
 \\
 4 - 3 x_{i1} \; , & \mbox{for} \; m_{i} \gg M_{1}
 \end{array} \; , 
\end{equation}
where $x_{ij} \equiv m_{i}^{2}/M_{j}^{2}$. As $M_{1} = M_{W} \ll M_{2}$, the term with $a=1$ in the summation dominates. Summing over all the $j$ gives the 
$(ee)$ element of the neutrino Dirac mass matrix,  $\sum_{j=1}^{6} m_{j} (V_{L})_{1j}(V_{R}^{\ast})_{1j} = (M_{\nu_{D}})_{ee}$. The electron EDM is then given by,
\begin{eqnarray}
d_{e} & \simeq & -\frac{e\alpha}{4\pi M_{W}^{2}} \frac{\kappa\kappa^{\prime}}{v_{R}^{2}-v_{L}^{2}} \mbox{Im}(e^{-i\alpha_{\kappa^{\prime}}}M_{D})_{ee} 
\\
& \simeq & 
 10^{-19} \times r \biggl( \frac{GeV}{v_{R}}\biggr)^{2} \biggl(\frac{ |(M_{\nu_{D}})_{ee} | }{MeV}\biggr) \sin(2\alpha_{\kappa^{\prime}}) \quad \mbox{e-cm} \; . \nonumber
\end{eqnarray}

In order to have a prediction for the electron EDM accessible to the next generation of experiments, $d_{e} \lesssim  10^{-32}$ e-cm~\cite{Lamoreaux:2001hb}, the $SU(2)_{R}$ breaking scale $v_{R}$ thus cannot be much higher than the weak scale. For the $(ee)$ element of the neutrino Dirac mass term $|(M_{\nu_{D}})_{ee}|$ of the order of the electron mass $\sim 1$ MeV, the required scale for $v_{R}$ has to be around $\sim \mathcal{O}(10^{5-6} \; \mbox{GeV})$. Such a low $SU(2)_{R}$ breaking scale can be obtained with the following $U(1)_{H}$ charge assignments:
\begin{equation}
Q(\Phi)  =  -Q(\tilde{\Phi}) = 2
\; , \; 
Q(\Delta_{L}) =  -Q(\Delta_{R}) = 4
\; , \; 
Q(L_{L})  =  -Q(L_{R}) = - 2
\; ,
\end{equation}
and $r = \omega^{2}$ with $\omega = 0.1$. These charge assignments lead to the following dominant contributions to the mass matrices,
\begin{eqnarray} 
v_{L} v_{R} & \simeq &  \beta^{\prime} \kappa^{2} \omega^{8}
\; , \\
M_{e} & \simeq & R_{ij} \kappa  \omega^{2} + \mathcal{O}(\omega^{6})
\; , \label{eq:me}\\
M_{\nu_{D}} & \simeq & R_{ij} \kappa e^{-i\alpha_{\kappa^{\prime}}} \omega^{4} + \mathcal{O}(\omega^{6})
\; , \\
M_{\nu}^{LL} & = & f_{ij} v_{L} e^{i\alpha_{L}} \; , \quad M_{\nu}^{RR} = f_{ij} v_{R} \; . 
\end{eqnarray}
The VEV $v_{L}$ which is compatible with the experimental observed mass squared differences in neutrino oscillation is of the order of $\sim (0.01 - 0.1)$ eV. This scale can be generated with $v_{R} \sim 10^{6}$ GeV. The effective neutrino mass matrix is then given by,
\begin{eqnarray}
M_{\nu}^{eff} & = & M_{\nu}^{LL} - M_{\nu_{D}} (M_{\nu}^{RR})^{-1} M_{\nu_{D}}^{T} \\ 
& = & v_{L} \bigl[ f_{ij} e^{i\alpha_{L}} - s R_{ij} f_{ij}^{-1} R_{ij}^{T}  e^{- 2 i \alpha_{\kappa^{\prime}}}
 \bigr] \; . \nonumber
\end{eqnarray}
where $s$ is the proportionality constant $1/\beta^{\prime}$~\cite{Chen:2004ww}. 
Under these charge assignments, the terms proportional to $R_{ij}$ in the charged lepton mass matrix and the neutrino Dirac mass matrix dominates. Thus the Yukawa coupling for the neutrino Dirac mass matrix is determined once the charged lepton mass matrix is known. This also allows the connection between CP violation in the quark sector, which is dictated solely by the phase $\alpha_{\kappa^{\prime}}$, and CP violation in the lepton sector.  The leptonic mixing matrix is obtained by diagonalizing the effective neutrino mass matrix, $M_{\nu}^{eff}$. As both $\alpha_{\kappa^{\prime}}$ and $\alpha_{L}$ now appear in $M_{\nu}^{eff}$, the leptonic CP phases, $\delta$, $\alpha_{12}$ and $\alpha_{13}$, as defined in Ref.~\cite{Chen:2004ww}, are thus functions of both of these two phases.

\section{Results}
\label{sec:result}

We assume that all the leptonic mixing results from the neutrino sector, and thus,
\begin{equation}
R_{ij} = \left(\begin{array}{ccc}
m_{e}/m_{\tau} & 0 & 0\\
0 & m_{\mu}/m_{\tau} & 0\\
0 & 0 & 1\end{array}\right) \; ,
\end{equation}
where 
the charged lepton masses roughly obeys $m_{e} : m_{\mu} : m_{\tau} = (\epsilon_{c})^{4} : (\epsilon_{c})^{2} : 1$, with the small parameter $\epsilon_{c}$ being roughly the size of the Cabbibo angle, $\epsilon_{c}=0.22$. The mass $m_{\tau}$ of order of $1$ GeV can thus be naturally generated with $R_{33} = 1$ due to the suppression factor $\omega^{2}$ in Eq.~\ref{eq:me}. The mixing pattern strongly depends on the flavor structure in the Yukawa matrix, $f_{ij}$ for the Majorana mass terms. It is taken to be,
\begin{equation}
f_{ij} = \left(
\begin{array}{ccc}
t^{2} & t & -t\\
t & 1 & 1\\
-t & 1 & 1
\end{array}\right) \; , 
\end{equation}
which only depends on one parameter, $t$, with $t < 1$ for the normal mass hierarchy among the light neutrinos in the low energy spectrum and gives rise to bi-large mixing pattern~\cite{Chen:2004ww}. 
The leptonic Jarlskog invariant in this case depends on the combination of CP phases, $(\alpha_{L} + \alpha_{\kappa^{\prime}})$,
\begin{equation}
J_{CP} = -\frac{(1-t^{2}) \sin(\alpha_{L} + \alpha_{\kappa^{\prime}})}{\Delta m_{21}^{2} \Delta m_{31}^{2} \Delta m_{32}^{2}} \biggl[
2 st^{2} \biggl(\frac{m_{e}}{m_{\tau}}\biggr) 
-s^{2}\biggl(\frac{m_{e}}{m_{\tau}}\biggr)^{2} \cos(\alpha_{L} + \alpha_{\kappa^{\prime}}) 
+ \frac{1}{8} \frac{s^{3}}{t^{2}} \biggl(\frac{m_{e}}{m_{\tau}}\biggr)^{3} \biggr] \; ,
\end{equation}
where $\Delta m_{ij}^{2} \equiv m_{i}^{2} - m_{j}^{2}$ (for $i, \, j = 1, \, 2, \, 3$) are the square mass differences  among the three light neutrinos.

\begin{figure}
\begin{tabular}{ccccc}
\includegraphics[scale=0.8]{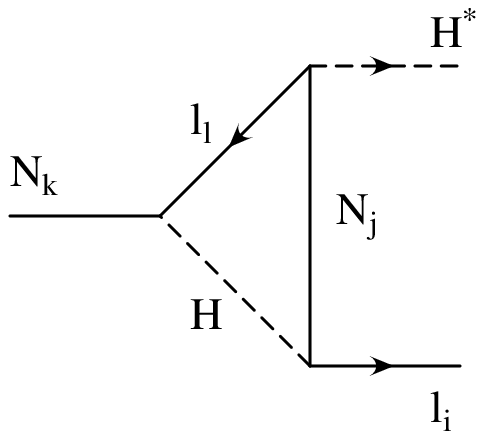} & 
\hspace{0.2in} & 
\includegraphics[scale=0.8]{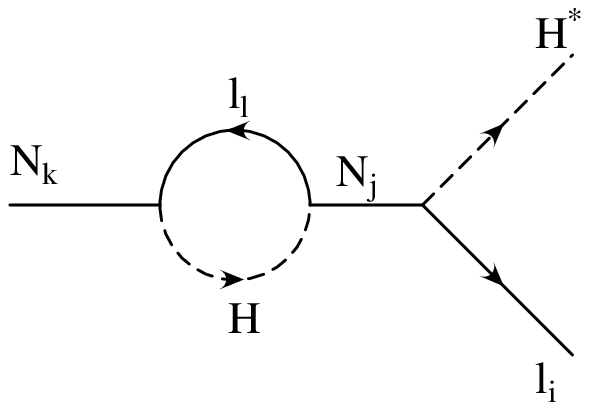} & 
\hspace{0.2in} & 
\includegraphics[scale=0.8]{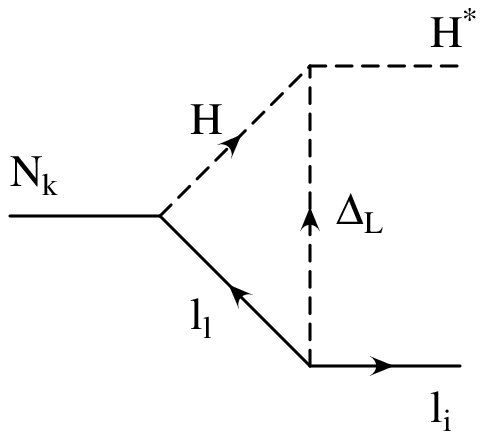} \\
    (a) &  & (b) &  & (c) \\
    \end{tabular}
    \caption{Diagrams in the minimal left-right model that contribute to the lepton number asymmetry through the decay of the RH neutrinos. }
    \label{fig:lepg}
\end{figure}

The lepton number asymmetry in this model can arise either through the decay of the lightest RH neutrino, or through the decay of the $SU(2)_{L}$ triplet Higgs, $\Delta_{L}$~\cite{Joshipura:2001ya}. 
Assuming the $SU(2)_{L}$ triplet Higgs is heavier than the lightest RH neutrino, the decay of the lightest RH neutrino dominates. With the particle spectrum of this model, there are three one-loop diagrams, as shown in Fig.~\ref{fig:lepg},
that contribute to the total lepton number asymmetry, $\epsilon$. 
The total asymmetry is given by,
\begin{equation}
\epsilon =  \epsilon^{N_{1}} + \epsilon^{\Delta_{L}} \; ,
\end{equation}
where $\epsilon^{N_{1}}$ is the contributions from diagrams that involve the Higgs doublet and charged lepton (Fig.~\ref{fig:lepg}a, ~\ref{fig:lepg}b), while $\epsilon^{\Delta_{L}}$ is the contribution of the diagram that involving the $SU(2)_{L}$ triplet Higgs (Fig.~\ref{fig:lepg}c),
\begin{eqnarray}
\epsilon^{N_{1}} & = & \frac{3}{16\pi} \biggl( \frac{M_{R_{1}}}{\kappa^{2}} \biggr)
\cdot 
\frac{\mbox{Im}[\mathcal{M}_{D} (M_{\nu}^{I})^{\ast} \mathcal{M}_{D}^{T}]_{11}}{(\mathcal{M}_{D}\mathcal{M}_{D}^{\dagger})_{11}} \; ,
\\
\epsilon^{\Delta_{L}} & = & 
\frac{3}{16\pi} \biggl(\frac{M_{R_{1}}}{\kappa^{2}}\biggr) 
\cdot \frac{\mbox{Im}[ \mathcal{M}_{D}(M_{\nu}^{II})^{\ast} \mathcal{M}_{D}^{T}]_{11}}{(\mathcal{M}_{D}\mathcal{M}_{D}^{\dagger})_{11}} \; .
\end{eqnarray}
where $\kappa \simeq v = 174 \; \mbox{eV}$ is the VEV of the $Y=+1$ component of the $\Phi$ field. 
Here $\mathcal{M}_{D}$ is the neutrino Dirac mass matrix in the basis in which RH neutrino mass matrix is real and diagonal. The matrices $M_{\nu}^{I} \equiv M_{\nu_{D}} (M_{\nu}^{RR})^{-1} M_{\nu_{D}}^{T}$ and $M_{\nu}^{II} \equiv M_{\nu}^{LL}$ are  the usual Type-I seesaw term and the LH Majorana mass term due to the $SU(2)_{L}$ triplet VEV, respectively.   As pointed out in Ref.~\cite{Chen:2004ww}, the only non-vanishing contribution is the one that involves the 
$SU(2)_{L}$ triplet in the loop, $\epsilon^{\Delta_{L}}$. This statement is still valid in the present case with complex neutrino Dirac mass matrix $M_{\nu_{D}} \propto e^{-i\alpha_{\kappa^{\prime}}}$, because the argument of $(M_{\nu}^{I})^{\ast} \propto e^{2i\alpha_{\kappa^{\prime}}}$ cancels the argument of $\mathcal{M}_{D} \mathcal{M}_{D}^{T} \propto e^{-2i\alpha_{\kappa^{\prime}}}$,  given that the RH Majorana mass matrix $M_{\nu}^{RR}$ is real. 
This is therefore a generic feature of the minimal LR model with SCPV. In Ref.~\cite{Chen:2004ww}, the contribution to the total asymmetry 
$\epsilon = \epsilon^{\Delta_{L}}$ is proportional to $\sim \sin\alpha_{L}$.  In the present case, the total asymmetry is proportional to 
$\sin(\alpha_{L} + 2 \alpha_{\kappa^{\prime}})$ as the Dirac mass matrix $M_{\nu_{D}}$ is now complex and proportional to $e^{i\alpha_{\kappa^{\prime}}}$. The total lepton number asymmetry $\epsilon$ is proportional to $\Delta \epsilon^{\prime}$ defined as~\cite{Chen:2004ww},
\begin{equation}
\Delta \epsilon^{\prime} = \frac{3f_{1}^{0}}{16\pi}  \frac{\mbox{Im} \bigl[\mathcal{M}_{D} ( f e^{i\alpha_{L}})^{\ast} \mathcal{M}_{D}^{T} \bigr]_{11}}{(\mathcal{M}_{D}\mathcal{M}_{D}^{\dagger})_{11}} = \frac{\epsilon^{\Delta_{L}}}{\beta} \, \propto \, \sin(\alpha_{L} + 2 \alpha_{\kappa^{\prime}}) \; ,
\end{equation}
where $f_{1}^{0}$ is the smallest eigenvalue of the matrix $f$. It is interesting to note that if $\alpha_{L} = -2 \alpha_{\kappa^{\prime}}$, the lepton number asymmetry vanishes, but $J_{CP}$ is non-zero. However, such a relation is generally not protected by any symmetry and will not hold when radiative corrections are taken into account. The total lepton asymmetry $\epsilon = \epsilon^{\Delta_{L}}$ is related to the 
observed baryon asymmetry of the universe, given in terms of the ratio of the baryon number $n_{b}$ to entropy $s$, as, 
\begin{equation}
\frac{n_{b}}{s} \simeq - \frac{24+4n_{H}}{66+13n_{H}} \, \epsilon \, \eta \, Y_{N_{1}}^{\mbox{eq}}(T \gg M_{1}) \; ,
\label{eq:asymmetry}
\end{equation}
where $Y_{N_{1}}^{\mbox{eq}}(T \gg M_{1}) = 135\zeta(3)/(4\pi^{4}g_{\ast})$ with $g_{\ast}$ being the number of spin-degrees of freedom in thermal equilibrium. The coefficient $(24+4n_{H})/(66+13n_{H})$, where $n_{H}$ is the number of Higgs doublets, is the fraction of the $B-L$ asymmetry that is converted into the baryon number asymmetry. For $g_{\ast} = 106.75$ as in the SM case and $n_{H} = 2$, the final asymmetry is given by, $n_{b}/s \simeq -1.38 \times 10^{-3} \, \epsilon \, \eta$.  
The efficiency factor $\eta$ measures the the number density of the right-handed neutrinos with respect to the equilibrium value, the out-of-equilibrium condition at the decay as well as the thermal corrections to the asymmetry. 
 It is obtained by solving the Boltzmann equations, and depends on the effective mass,
\begin{equation}
\widetilde{m} \equiv \frac{(\mathcal{M}_{D}\mathcal{M}_{D}^{\dagger})_{11}}{M_{1}} \; ,
\end{equation}
as well as the mass of the lightest right-handed neutrino, $M_{1}$.  
The BBN and WMAP measurements imply that  $n_{b}/s = (0.87 \pm 0.04) \times 10^{-10}$, assuming standard $\Lambda$CDM cosmological model~\cite{Bennett:2003bz}.

We search the allowed parameter space for $(t,s,\alpha_{\kappa^{\prime}}, \alpha_{L})$ that satisfy the $1\sigma$ 
limits for the oscillation parameters~\cite{Maltoni:2004ei}: $\Delta m_{23}^{2} = (2.4 \pm 0.3) \times 10^{-3} \; \mbox{eV}^{2}$, 
$\Delta m_{12}^{2} = (7.9 \pm 0.4) \times 10^{-5} \; \mbox{eV}^{2}$, $\sin^{2} 2\theta_{23} > 0.9 $, and $0.4 < \tan^{2} \theta_{12} < 0.509$. Using the allowed parameter space, the predictions for various CP violating processes can then be made. In Fig.~\ref{fig:edmlpg}, the correlation between the electron EDM and the amount of lepton number asymmetry $\Delta \epsilon^{\prime}$ is shown. For $\alpha_{L}= 0$, all CP violation in both the quark and lepton sectors depends on only one phase, $\alpha_{\kappa^{\prime}}$. One thus find a very strong correlation between $d_{e}$ and $\epsilon$ as these two physical observables are proportional to each other. In Fig.~\ref{fig:jcplpg} the $J_{CP}$ dependence on the lepton number asymmetry, $\epsilon$, and in the electron EDM, $d_{e}$,  is shown. For $\alpha_{L}=0$, one finds a reflection symmetry between the first and the forth quadrants as $\epsilon$, $J_{CP}$ and $d_{e}$ are proportional to $\sin\alpha_{\kappa^{\prime}}$ in this case. In Fig.~\ref{fig:jcpnulb}, the correlation between the leptonic Jarlskog $J_{CP}$ and the matrix element for neutrinoless double beta decay $\left< m_{0\nu\beta\beta}\right>_{ee}$ is shown. It is found that the correlation is independent of whether $\alpha_{L}$ vanishes or not.  This exhibits the fact that  the neutrinoless double beta decay depends on the the Majorana phases while $J_{CP}$ depends on the Dirac phase and that all three low energy leptonic CP phases $\delta$, $\alpha_{12}$ and $\alpha_{13}$ ranging from $0$ to $2\pi$ can be generated irrespective of whether $\alpha_{\kappa^{\prime}}$ vanishes or not. In Fig.~\ref{fig:ue3lpg}, the correlation between the lepton number asymmetry and the $|U_{e3}|$ element in the leptonic mixing matrix is given. The prediction for $|U_{e3}|$ ranges from $0.01$ to the current upper bound, $0.16$, even though most of the allowed parameter space for $(t,\, s, \, \alpha_{\kappa^{\prime}}, \, \alpha_{L})$ gives rise to predictions that range between $0.01$ and $0.03$. Fig.~\ref{fig:blpg0} shows the lepton number asymmetry $\epsilon$ as a function of the CP phase $\alpha_{\kappa^{\prime}}$. The total lepton asymmetry in our model is  $\epsilon^{\Delta_{L}} = \beta \Delta \epsilon^{\prime} \sim \mathcal{O}(10^{-9})$, taking into account the suppression factor $\beta \sim \omega^{8}$.  
For the decay of the lightest right-handed neutrino to be out-of-equilibrium, the effective mass for leptogenesis is required to be,
$\widetilde{m}  \lesssim \mathcal{O}(0.01) \; \mbox{eV}$. 
As the allowed parameter space gives rise to an effective mass $\widetilde{m}$ in the range $(0.0007 \sim 0.018)$ eV, the washout effects are negligible in this model.  Furthermore, the efficiency factor $\eta$ given in Eq.~\ref{eq:asymmetry} only depends on the effective mass, $\widetilde{m}$, as $M_{1}$ in our model is much smaller than $10^{14}$ GeV~\cite{Giudice:2003jh}. For $\widetilde{m} \sim (10^{-4}-10^{-3} ) \; \mbox{eV}$, $\eta$ can be as large as $\mathcal{O}(10^{1}-10^{2})$, provided that the right-handed neutrinos abound dominantly at early times~\cite{Giudice:2003jh}. This thus allows sufficient baryon number asymmetry to be generated in our model even with $M_{R}$ of $\sim10^{6}$ GeV.

\section{Conclusion}
\label{sec:conclude}

It is in general not possible to relate leptogenesis to low energy leptonic CP violation processes due to the present of additional mixing angles and CP phases in the heavy neutrino sector. We found that, in minimal left-right symmetry model with spontaneous CP violation, such connection can be established, as there are only two physical phases in this model. With the additional $U(1)_{H}$ horizontal symmetry, the left-right breaking scale $v_{R}$ can be lowered to $10^{3}$ TeV, thus allowing observable EDM of the electron. Due to this $U(1)_{H}$ symmetry, the phase $\alpha_{\kappa^{\prime}}$ that governs all CP violation in the quark sector can now be relevant for the lepton sector, thus giving rise to relations between CP violating processes in the two sectors. There, however, exist two serious issues in the minimal left-right model with spontaneous CP violation. One is the fine-tuning required to generate sizable $\alpha_{\kappa^{\prime}}$. We find that, this fine-tuning problem cannot be solved by lowering the left-right breaking scale $v_{R}$ with an additional $U(1)_{H}$ symmetry. And one may need to call for extra dimensions or by introducing additional scalar fields for spontaneous CP violation to work in the quark sector. It should be noted, however, that there is no such constraint in the lepton sector. Secondly, it has been pointed out previously that $\alpha_{\kappa^{\prime}}$  alone cannot give rise to sufficient CP violation to account for the observed asymmetry in $B \rightarrow \psi K_{s}$ decay. It 
would be interesting to see if this could be resolved in the supersymmetric case. These issues are under investigation.


\begin{acknowledgments}
We thank Gabriela Barenboim and Uli Nierste for discussion. Fermilab is operated by Universities Research Association Inc. under Contract No. DE-AC02-76CH03000 with the U.S. Department of Energy. The work of KTM was supported, in part, by the Department of Energy under Grant no. 
DE-FG03-95ER40892. M-CC acknowledges Aspen Center for Physics, where part this work was done, for its hospitality and for providing a intellectually stimulating atmosphere. 
\end{acknowledgments}

\appendix

\section{Scalar Potential}

The complete scalar potential of this model is given by~\cite{Gunion:1989in,Deshpande:1990ip} 
\begin{equation}
V =  V_{\Phi} +V_{\Delta} +V_{\Phi \Delta} \; ,
\end{equation}
where
\begin{eqnarray}
V_{\Phi} & = & - \mu^{2} \mbox{Tr}(\Phi^{\dagger} \Phi) - \mu_{2}^{2} \bigl[ \mbox{Tr}(\tilde{\Phi}\Phi^{\dagger}) + \mbox{Tr}(\tilde{\Phi}^{\dagger}\Phi)\bigr]
\\
&&
+ \lambda_{1} \bigl[ \mbox{Tr} (\Phi\Phi^{\dagger}) \bigr]^{2} + \lambda_{2} \biggl[  \bigl[\mbox{Tr}(\tilde{\Phi}\Phi^{\dagger})\bigr]^{2}
+ \mbox{Tr} ( \tilde{\Phi}^{\dagger} \Phi) \bigr]^{2} \biggr] \nonumber\\
& & + \lambda_{3} \bigl[ \mbox{Tr}(\tilde{\Phi}\Phi^{\dagger}) \mbox{Tr}(\tilde{\Phi}^{\dagger} \Phi) \bigr] 
+ \lambda_{4} \biggl[
\mbox{Tr}(\Phi\Phi^{\dagger}) \bigl[ \mbox{Tr} (\tilde{\Phi} \Phi^{\dagger}) + \mbox{Tr}(\tilde{\Phi}^{\dagger} \Phi) \bigr]\biggr] \; ,
\nonumber
\end{eqnarray}
\begin{eqnarray}
V_{\Delta} & = & -\mu_{3}^{2} \bigl[ \mbox{Tr}(\Delta_{L} \Delta_{L}^{\dagger}) + \Delta_{R} \Delta_{R}^{\dagger})\bigr]
+ \rho_{1} \biggl[
\bigl[\mbox{Tr}(\Delta_{L}\Delta_{L}^{\dagger}) \bigr]^{2} + \bigl[\mbox{Tr}(\Delta_{R}\Delta_{R}^{\dagger})\bigr]^{2}\biggr]
\\
&& +  \rho_{2} \biggl[
\mbox{Tr}(\Delta_{L}\Delta_{L}) \mbox{Tr}(\Delta_{L}^{\dagger}\Delta_{L}^{\dagger})
+ \mbox{Tr}(\Delta_{R}\Delta_{R}) \mbox{Tr}(\Delta_{R}^{\dagger}\Delta_{R}^{\dagger})\biggr]
\nonumber\\
&& +  \rho_{3} \bigl[
\mbox{Tr}(\Delta_{L}\Delta_{L}^{\dagger}) \mbox{Tr}(\Delta_{R}\Delta_{R}^{\dagger})\bigr]
+ \rho_{4}\biggl[
\mbox{Tr}(\Delta_{L}\Delta_{L}) \mbox{Tr}(\Delta_{R}^{\dagger}\Delta_{R}^{\dagger})
+\mbox{Tr}(\Delta_{L}^{\dagger}\Delta_{L}^{\dagger}) \mbox{Tr}(\Delta_{R}\Delta_{R}) \biggr] \; ,
\nonumber
\end{eqnarray}
\begin{eqnarray}
V_{\Phi\Delta} & = & \alpha_{1}\biggl[  \mbox{Tr}(\Phi\Phi^{\dagger}) 
\bigl[ \mbox{Tr}(\Delta_{L}\Delta_{L}^{\dagger}) + \mbox{Tr}(\Delta_{R}\Delta_{R}^{\dagger}) \bigr] \biggr]  
\\
&& + \alpha_{2}\bigl[  \mbox{Tr}(\Phi \tilde{\Phi}^{\dagger})  \mbox{Tr}(\Delta_{R}\Delta_{R}^{\dagger}) 
+  \mbox{Tr}(\Phi^{\dagger} \tilde{\Phi})  \mbox{Tr}(\Delta_{L}\Delta_{L}^{\dagger}) \bigr] \biggr]   
\nonumber\\
&& + 
\alpha_{2}^{\ast} \bigl[  \mbox{Tr}(\Phi^{\dagger} \tilde{\Phi})  \mbox{Tr}(\Delta_{R}\Delta_{R}^{\dagger}) 
+  \mbox{Tr}(\tilde{\Phi}^{\dagger} \Phi)  \mbox{Tr}(\Delta_{L}\Delta_{L}^{\dagger}) \bigr] \biggr] 
\nonumber\\
&& + \alpha_{3} \bigl[  \mbox{Tr}(\Phi \Phi^{\dagger} \Delta_{L}\Delta_{L}^{\dagger}) 
+  \mbox{Tr}(\Phi^{\dagger} \Phi\Delta_{R}\Delta_{R}^{\dagger}) \bigr] 
+ \beta_{1} \bigl[ 
\mbox{Tr}( \Phi \Delta_{R} \Phi^{\dagger} \Delta_{L}^{\dagger} )
+ \mbox{Tr}(\Phi^{\dagger} \Delta_{L} \Phi \Delta_{R}^{\dagger}) \bigr]
\nonumber\\
&& + \beta_{2} \bigl[ 
\mbox{Tr}( \tilde{\Phi} \Delta_{R} \Phi^{\dagger} \Delta_{L}^{\dagger} )
+ \mbox{Tr}(\tilde{\Phi}^{\dagger} \Delta_{L} \Phi \Delta_{R}^{\dagger}) \bigr]
+ \beta_{3} \bigl[ 
\mbox{Tr}( \Phi \Delta_{R} \tilde{\Phi}^{\dagger} \Delta_{L}^{\dagger} )
+ \mbox{Tr}(\Phi^{\dagger} \Delta_{L} \tilde{\Phi} \Delta_{R}^{\dagger}) \bigr] \; .
\nonumber
\end{eqnarray}
Minimization of this scalar potential leads to the following relations,
\begin{eqnarray}
(2\rho_{1}-\rho_{3}) v_{R} v_{L} & = &  \beta_{1} \kappa \kappa^{\prime} \cos(\alpha_{L} - \alpha_{\kappa^{\prime}}) + \beta_{2} \kappa^{2} \cos\alpha_{L} 
+ \beta_{3} \kappa^{\prime \, 2} \cos(\alpha_{L} - 2\alpha_{\kappa^{\prime}}) \; ,
\label{eq:minimize1}
\\
 0 & = & \beta_{1} \kappa \kappa^{\prime} \sin(\alpha_{L} - \alpha_{\kappa^{\prime}}) + \beta_{2} \kappa^{2} \sin\alpha_{L} 
+ \beta_{3} \kappa^{\prime \, 2} \sin(\alpha_{L}-2\alpha_{\kappa^{\prime}}) \; ,
\label{eq:minimize2}
\\
 0 & = & v_{R} v_{L} \biggl[ 2 \kappa\kappa^{\prime} (\beta_{2} + \beta_{3}) \sin(\alpha_{L} - \alpha_{\kappa^{\prime}}) + \beta_{1}\bigl[
\kappa^{2} \sin\alpha_{L} + \kappa^{\prime \, 2} \sin(\alpha_{L} - 2 \alpha_{\kappa^{\prime}}) \bigr]  \biggr]
\nonumber\\
& &   + \kappa\kappa^{\prime} \sin\alpha_{\kappa^{\prime}} \bigl[
\alpha_{3} (v_{L}^{2} + v_{R}^{2}) + (4\lambda_{3} - 8\lambda_{2}) (\kappa^{2} - \kappa^{\prime \, 2})\bigr] \; .
\label{eq:minimize3}
\end{eqnarray}


\bibliography{leptog,edm,lr,nuexp,so10,ed,scpv}




\begin{figure}[h!]
\begin{center}
\includegraphics[scale=0.4]{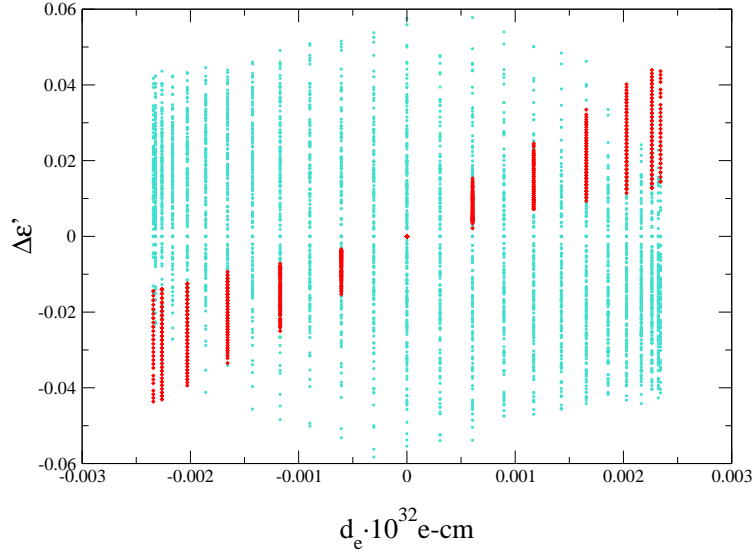}
\caption{The electron EDM $d_{e}$  versus lepton number asymmetry $\Delta \epsilon^{\prime}$. The red dots (darker shade) correspond to $\alpha_{L}=0$.}
\label{fig:edmlpg}
\end{center}
\end{figure}

\begin{figure}[h!]
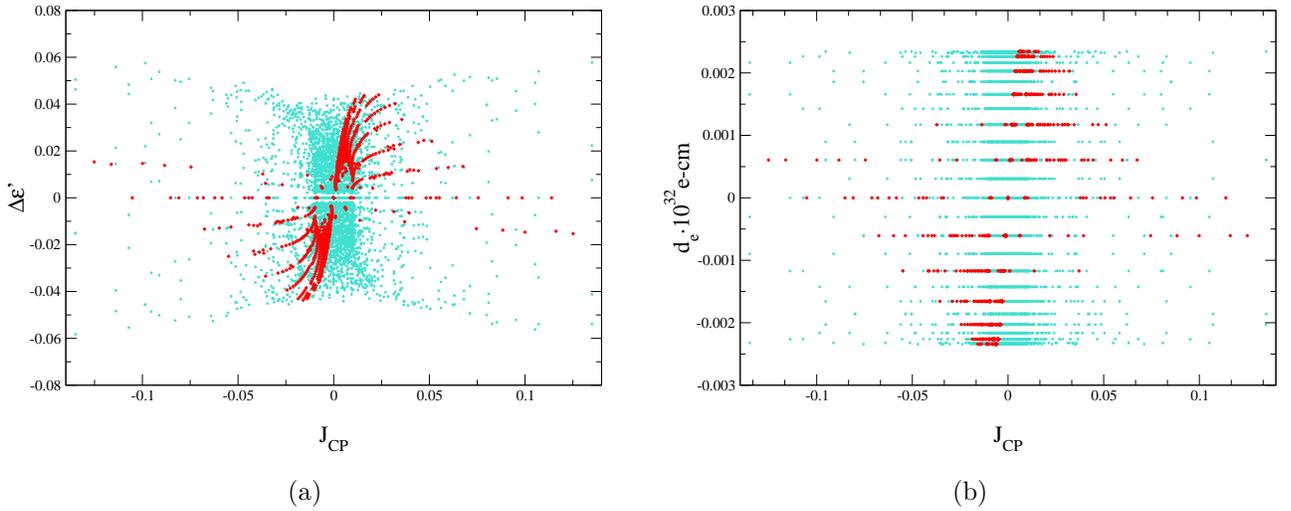

\begin{tabular}{ccc}
\includegraphics[scale=0.33]{jcp-lpg.eps} & \hspace{0.5cm} & 
\includegraphics[scale=0.33]{jcp-edm.eps}
\\
(a) & & (b)
\end{tabular}
\caption{(a) The dependence on the leptonic Jarlskog invariant $J_{CP}$ in lepton number asymmetry $\Delta \epsilon^{\prime}$. (b) 
The dependence on the leptonic Jarlskog invariant $J_{CP}$ in the elecrton EDM $d_{e}$. The red dots (darker shade) correspond to $\alpha_{L}=0$.}
\label{fig:jcplpg}
\end{figure}

\begin{figure}[h!]
\begin{center}
\includegraphics[scale=0.4]{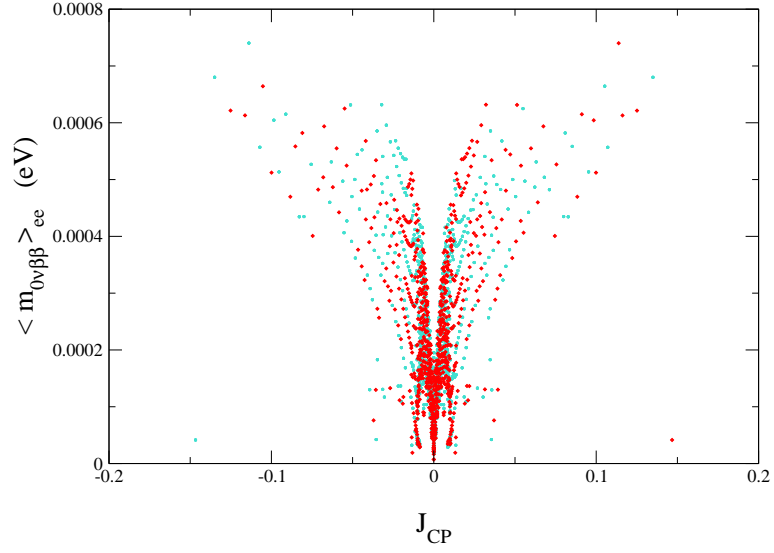}
\caption{Correlation between the leptonic Jarlskog $J_{CP}$ and matrix element of neutrinoless double beta decay $\left<m_{ee}\right>_{0\nu\beta\beta}$. The red dots (darker shade) correspond to $\alpha_{L}=0$.}
\label{fig:jcpnulb}
\end{center}
\end{figure}

\begin{figure}[h!]
\includegraphics[scale=0.4]{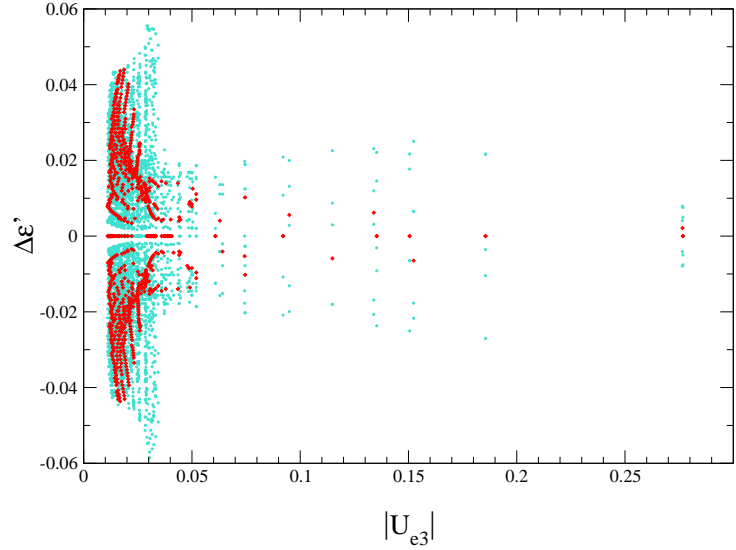} 
\caption{Correlation between the lepton number asymmetry $\Delta \epsilon^{\prime}$ and the element $|U_{e3}|$ in the leptonic mixing matrix.  The red dots (darker shade) correspond to $\alpha_{L}=0$.}
\label{fig:ue3lpg}
\end{figure}

\begin{figure}[h!]
\includegraphics[scale=0.33]{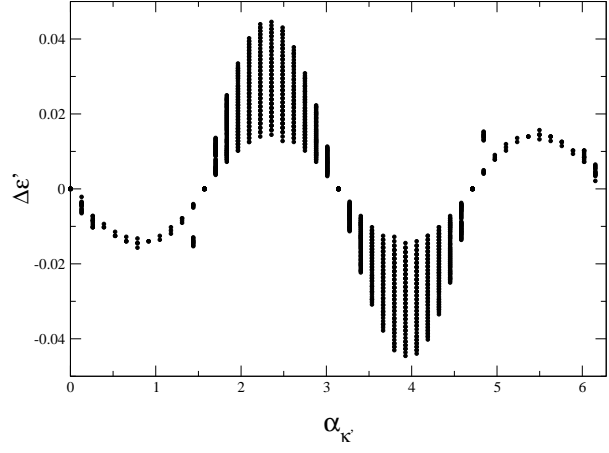}
\caption{The lepton number asymmetry as a function of the phase $\alpha_{\kappa^{\prime}}$ with $\alpha_{L}=0$.}
\label{fig:blpg0}
\end{figure}

\end{document}